\documentclass[aps,prb,twocolumn,preprintnumbers,amsmath,amssymb]{revtex4}
\usepackage{graphicx}
\usepackage{xcolor}
\begin{document}

\title{Spin-fluctuation heat capacity at magnetic phase transition in the Co,Fe doped MnSi.}
\author{S.~M.~Stishov}
\email{stishovsm@lebedev.ru}
\affiliation{P. N. Lebedev Physical Institute, Leninsky pr., 53, 119991 Moscow, Russia}
\author{A.~E.~Petrova}
\affiliation{P. N. Lebedev Physical Institute, Leninsky pr., 53, 119991 Moscow, Russia}
\author{A. M. Belemuk}
\affiliation{Institute for High Pressure Physics of RAS, 108840 Troitsk, Moscow, Russia}

\begin{abstract}
An universal line revealing an independence of spin fluctuation contributions to the heat capacity on impurity content and its nature is discovered in the helical phase of Mn(Co,Fe)Si. This situation  declares an invariance of the heat capacity of spin subsystem under doping, which   probably arises as a result of relative stiffness of the helical spin structure in respect to the impurity spins. On the other hand the situation drastically changes at the helical fluctuation region when no long range spin order exists. 
\end{abstract}

\maketitle

\section{Introduction}

\begin{figure}[htb]
\includegraphics[width=80mm]{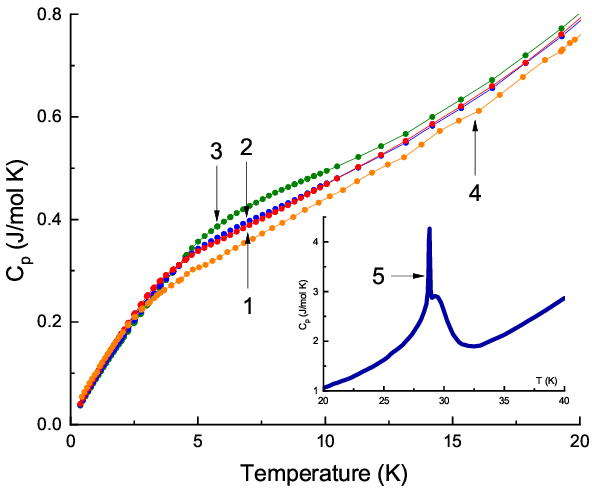}
\caption{\label{fig1} Heat capacity of Mn$_{1-x}$(Co,Fe)$_{x}$Si in the transition region from helical to paramagnetic structure. T (1–4: x = 0.057 (Co), 0.063(Co), 0.09(Co), 0.17 (Fe), 5-MnSi ). After Ref.~\cite{Pet2,Pet3,Pet1,stpt}.} 
\end{figure}

\begin{figure}[htb]
\includegraphics[width=60mm]{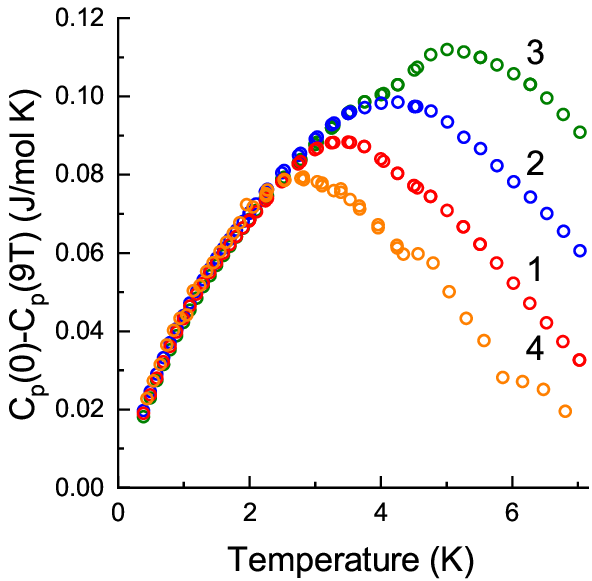}
\caption{\label{fig2} Difference between heat capacity at zero magnetic field $C_p(0)$ and heat capacity at 9~T ($C_p(9 T)$ for (Mn,Co)Si samples  and the sample of (Mn,Fe)Si, revealing strange scaling behavior at low temperatures.  As seen the Co and Fe doping destroys the first-order phase transition peak, spreads out fluctuation maxima, and shifts maxima to lower  temperatures.(1–4: x = 0.057 (Co), 0.063(Co), 0.09(Co), 0.17 (Fe)). After Ref.~\cite{Pet2,Pet3,Pet1}.} 
\end{figure}

In the course of recent study~\cite{Pet1} an unexpected  scaling behavior of heat capacity was seen (Figs~\ref{fig1}, \ref{fig2}), which served as a motivation for the current investigation. As is seen in Fig.~\ref{fig1} doping of MnSi with Co and Fe smears out the sharp phase transition (see the inset 5 in Fig.~\ref{fig1}) and makes it difficult to determine the transition area. A subtraction from the heat capacity curves at zero magnetic field the corresponding curves at 9 T as is shown in Fig.~\ref{fig2} helps to see clearly the transition and at the same time reveals the mantioned scaling behaviour of heat capacity. Note that this manipulation implies a subtraction of some background contributions, including
phonon and electron ones to the heat capacity leaving the spin fluctuation part intact. 
\begin{figure}[htb]
\includegraphics[width=60mm]{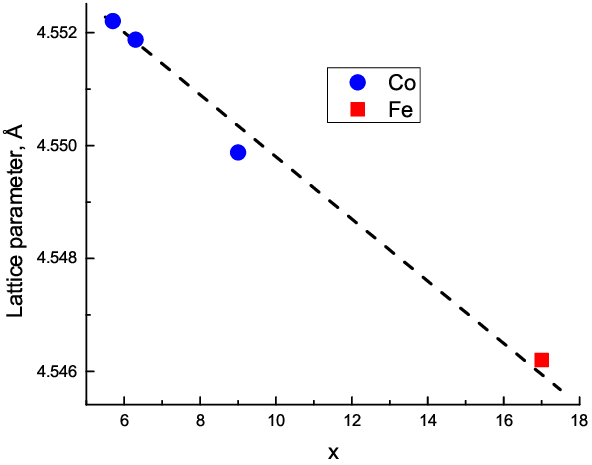}
\caption{\label{fig3}The lattice parameters of (Mn$_{1-x}$Co$_x$)Si: x=0.057,0.063,0.09 and (Mn$_{0.83}$Fe$_{0.17}$)Si Ref.~\cite{Pet1,Pet2,Pet3}.} 
\end{figure}
A preliminary analysis has shown that  physical properties of the Fe doped sample (Mn$_{0.83}$Fe$_{0.17}$)Si generally agree with a corresponding concentration dependence of properties of Co doped samples (see for instance Fig.~\ref{fig3}). This situation clearly demonstrates that a geometry factor (atomic radius of dopant) plays a major role in variation of physical properties of MnSi, which therefore are  results of volume changes at doping. The latter justifies our analysis of properties of Co and Fe doped samples as a single set of data thus expanding an available concentration range.                  
\section{Monte Carlo calculations and discussion}

Having tried to inderstand the scaling data the classical Monte Carlo (MC) calculations were made to describe the behavior of heat capacity of a spin system S$_{1-x}$I$_{x}$, where S-regular spin, I-impurity spin. We use an approach involving localized spins coupled by the exchange and Dzyloshinski-Moriya (DM) spin-spin interactions~\cite{Ono,Ham,Buh}. Upon doping a regular spin is replaced by an effective impurity spin  which is supposed to be a classical spin of unit length similar to the regular one. The impurity spins are coupled with neighboring regular spins by some modified exchange and DM coupling constants. If two impurity spins happen to occur in neighboring sites the corresponding coupling constants is forced to be zero. 

\begin{figure}[htb]
\includegraphics[width=80mm]{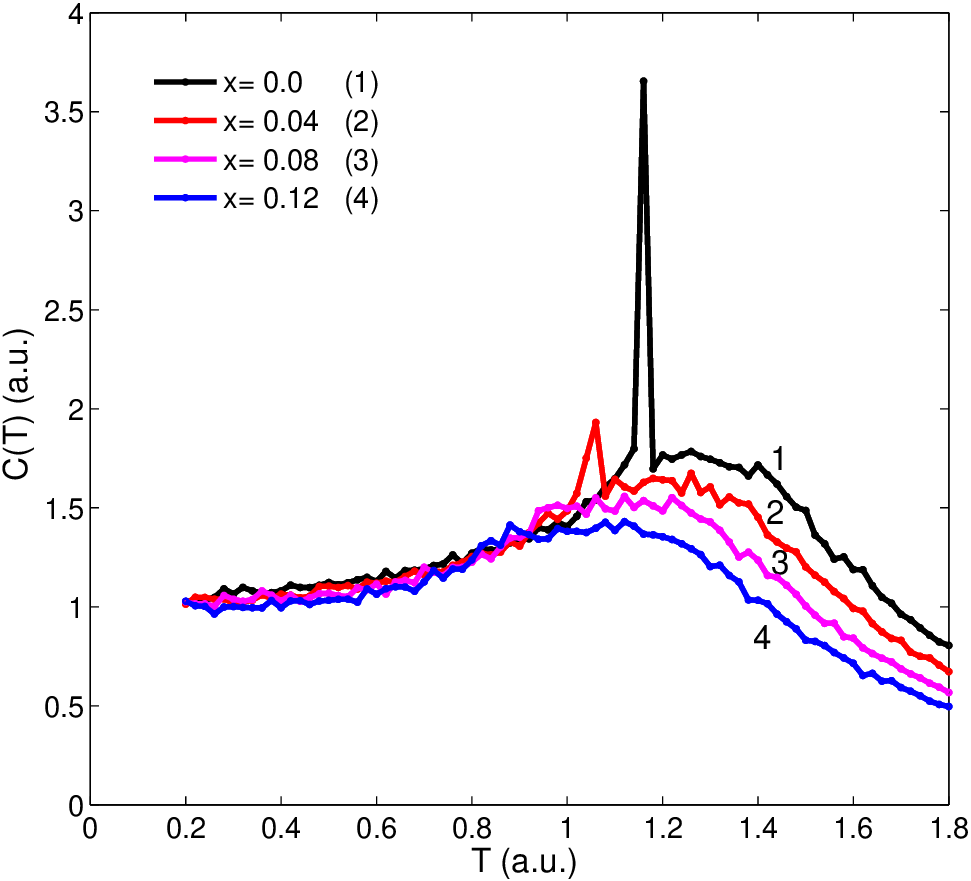}
\caption{\label{fig4} Monte Carlo calculations of heat capacity of classical spin system with impurities S$_{1-x}$I$_{x}$, showing different impurity influence on heat capacity of  the system in spin ordered and disordered phase. } 
\end{figure}

\begin{figure}[htb]
\includegraphics[width=80mm]{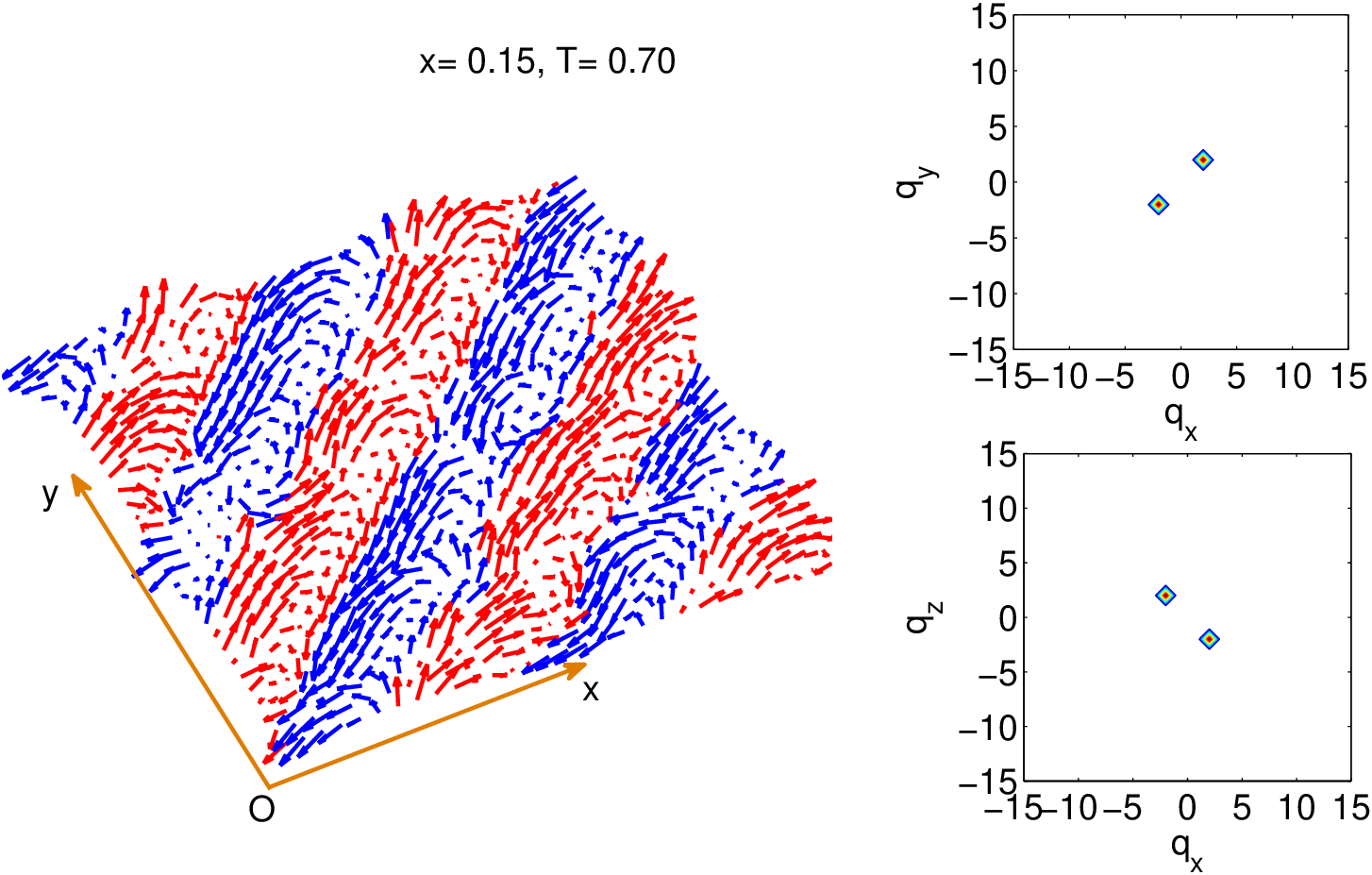}
\includegraphics[width=80mm]{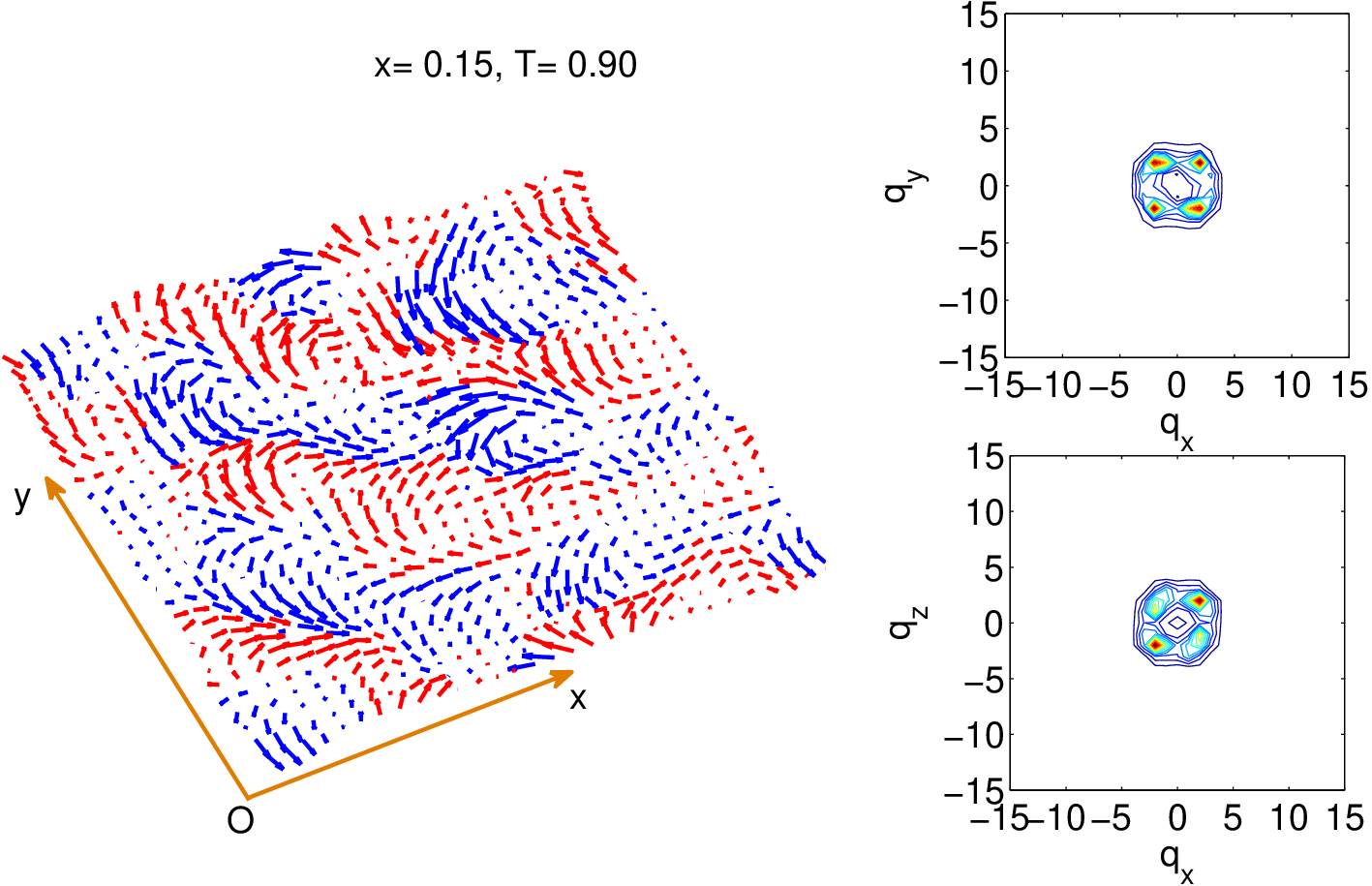}
\caption{\label{fig5} Spin configuration and profiles of the Bragg intensity for classical spin system S$_{1-x}$I$_{x}$ with 
doping concentration x = 0.15 before the transition T = 0.70 (upper) and after (bottom) transition T = 0.90. Spins with positive (negative) values of $S_{z}$ are presented in red (blue). The size of the arrows is proportional
to $\langle S_{i}\rangle$.} 
\end{figure}

The MC simulation was carried out on a  cubic lattice with periodic boundary conditions using a standard single-site Metropolis algorithm. We made $10^6$ MC steps per spin (MCS) to equilibrate the system and next $10^6$ MCS (and up to $10^7$ MCS in separate runs) to gain statistics.
The Monte Carlo heat capacity and spin structure calculations were also carried out in the transition region of the spin system with  isomorphic impurities, i.e. impurity spins replacing regular ones (see Figs.\ref{fig4},\ref{fig5}).  More details on the MC procedure with impurities and the choice of the coupling parameters of the effective model given in Refs. \cite{Bel1,Bel2}.


As is seen in Fig.~\ref{fig2} the mentioned procedure results in a puzzling universal line exposing an independence of the fluctuation contributions to the heat capacity on the impurity contents and its nature at temperatures below the transition region. This situation suggests an invariance of this contribution  despite the all changes caused by doping. At the same time the heat capacity lines of the substances with different impurities concentrations are significantly split at high temperatures in the paramagnetic phase with strong helical fluctuations, and where the heat capacity progressively decreases with an increase of impurity concentration. This is well mimicked by the Monte Carlo simulation, illustrated in Figs.~\ref{fig4} and \ref{fig5}. As is seen in Fig.~\ref{fig4} the heat capacity of the spin systems does not depend much on the impurity concentration at low temperatures region demonstrating the long range helical order (see Fig.~\ref{fig5}). Concurrently at high temperatures heat capacity the spin system  split in a number of branches with different impurity concentration showing the same trend like it occurs in the real system (Fig.~\ref{fig2}). Remarkable that this behavior of the heat capacity happens in the strong helical fluctuation region lacking a long range order.   	   

.

\section{Conclusion}
The mysterious universal line revealing an independence of spin fluctuation contributions to the heat capacity of Mn(Fe,Co)Si on impurity contents and its nature is discovered in the helical phase of Mn(Co,Fe).   This situation  declares an invariance of the heat capacity of spin subsystem under doping, which   probably arises as a result of relative stiffness of the helical spin structure in respect to the impurity spins.  Indeed, thermal excitation in a spin system with a long range order should be collective ones, so impurity spins cannot alter much its energy spectrum.  On the other hand the situation drastically changes at the helical fluctuation region when no long range spin order exists. Than thermal excitation certainly became localized and sensible to various kind of perturbations including impurity spins.    
However a complete explanation of this phenomena cannot be given at present time and it will be of a subject of a thorough study at immediate future. 
	
\end{document}